\documentclass[aps, pra, superscriptaddress, reprint, floatfix]{revtex4-2}

\usepackage{graphicx}
\usepackage[colorlinks=true, linkcolor=blue, citecolor=blue, urlcolor=blue]{hyperref}
\usepackage{amsmath}
\usepackage{amssymb}
\usepackage{amsfonts}
\usepackage{amsthm}
\usepackage{color}
\usepackage{physics}
\usepackage{booktabs}
\usepackage[capitalize]{cleveref}
\usepackage{etoolbox}
\usepackage{orcidlink}
\usepackage{multirow}
\makeatletter
\AddToHook{cmd/appendix/before}{\def\cref@section@alias{appendix}}
\makeatother

\AtBeginEnvironment{subequations}{%
}

\Crefname{equation}{Eq.}{Eqs.}
\Crefname{section}{Sec.}{Secs.}
\Crefname{figure}{Fig.}{Figs.}
\Crefname{figure}{Figure}{Figures}
\Crefname{appendix}{Appendix}{Appendices}
\crefname{table}{Table}{Tables}
\Crefname{table}{Table}{Tables}

\theoremstyle{plain}

\AfterEndEnvironment{theorem}{\noindent\ignorespaces}
\AfterEndEnvironment{lemma}{\noindent\ignorespaces}
\AfterEndEnvironment{corollary}{\noindent\ignorespaces}

\AfterEndEnvironment{definition}{\noindent\ignorespaces}
\AfterEndEnvironment{proof}{\noindent\ignorespaces}

\crefname{theorem}{Theorem}{theorems}
\crefname{lemma}{Lemma}{lemmas}
\crefname{corollary}{Corollary}{corollaries}
\crefname{definition}{Definition}{definitions}

\makeatletter
\renewcommand{\fnum@figure}{\textbf{\figurename~\thefigure}}
\makeatother

\begin{document}

\title{Shuttling-aware dynamical decoupling for quantum charge-coupled devices}

\author{Linus Schulte\,\orcidlink{0009-0002-0964-9087}}
\affiliation{Chair for Design Automation, Technical University of Munich, Munich, Germany}
\email{l.schulte@tum.de}

\author{Aaron Sander\,\orcidlink{0009-0007-9166-6113}}
\affiliation{Chair for Design Automation, Technical University of Munich, Munich, Germany}

\author{Robert Wille\,\orcidlink{0000-0002-4993-7860}}
\affiliation{Chair for Design Automation, Technical University of Munich, Munich, Germany}
\affiliation{MQSC GmbH, Munich, Germany}
\affiliation{Software Competence Center Hagenberg GmbH (SCCH), Hagenberg, Austria}

\date{\today}

\begin{abstract}
Dynamical decoupling (DD) helps maintain high-fidelity quantum computations by suppressing dephasing noise through carefully timed refocusing pulses. In quantum charge-coupled device (QCCD) architectures, however, where ions are shuttled throughout the device, transport constrains when pulses can be applied and affects the phase accumulated by an ion. Conventional DD methods do not account for shuttling and may therefore schedule pulses that must be omitted or shifted after transport scheduling, weakening the protection from dephasing.
We therefore introduce shuttling-aware dynamical decoupling (SADD), an offline compiler pass that jointly selects refocusing pulses and local ion rerouting while preserving logical-gate timings and the total schedule length. In benchmark simulations, SADD improves average final-state fidelity over both the original schedules and a simple nearest-feasible Hahn-echo baseline when dephasing dominates control and transport errors and varies slowly enough for DD. Rerouting enables otherwise infeasible pulse timings, while spatial information about the noise can provide further gains. These benefits disappear, however, when the added transport introduces too much error. Overall, our results show that coordinating DD with ion transport is an effective compiler strategy for reducing dephasing in QCCD processors.

\end{abstract}

\maketitle

\section{Introduction}

Trapped ions are a promising platform for quantum computing, but scaling beyond individual ion chains while preserving high-fidelity operation remains a challenge~\cite{Bruzewicz2019TrappedIons}. Quantum charge-coupled device (QCCD) architectures address the scaling problem by shuttling ions between localized processing zones, providing effective all-to-all connectivity through local gates~\cite{Kielpinski2002QCCD,Pino2021HoneywellQCCD}. At the same time, QCCD execution involves periods of transport and storage during which slowly varying dephasing can reduce coherence~\cite{Harty2014,Ruster2016longlived,Ransford2026Helios}.

A common method to suppress this dephasing is dynamical decoupling (DD), which applies carefully timed refocusing pulses~\cite{Viola1999Dynamical,Yang2011Preserving}, with existing compiler-level methods adapting pulse selection and timing to circuit structure, idle periods, and measured device behavior~\cite{Das2021ADAPT,Niu2022Analyzing,Niu2022Effects,Smith2022TimeStitch,Seif2024Suppressing,Coote2025ResourceEfficient}. However, these methods target architectures with stationary qubits and effectively ubiquitous local control, whereas in QCCD architectures, individually addressed control is restricted to designated processing zones and thus depends on the ions’ transport schedule. A pulse requested before these constraints are resolved may therefore be shifted or precluded by transport, logical operations, or processing-zone occupancy. Circuit-level DD optimization alone therefore does not ensure that the intended noise suppression survives physical QCCD scheduling.

Recent work by Watkins et al.\ demonstrates this failure mode directly on a Quantinuum QCCD processor~\cite{Watkins2026Improving}, where scheduling-induced offsets from nominal DD pulse times increase sensitivity to low-frequency noise. Compile-time timing feedback and opportunistic real-time pulse insertion can mitigate these offsets. Measurements on the 98-qubit Helios processor furthermore identify a schedule-dependent magnetic contribution to memory error during transport, suggesting a coupling between transport and dephasing exposure, with DD being proposed as a potential remedy~\cite{Ransford2026Helios}. A distinct offline compilation problem thus remains open, where we ask whether pulse placement and ion transport can be jointly optimized within a resolved QCCD schedule to realize effective dynamical decoupling against slowly varying dephasing noise.

Here, we introduce a modular schedule-augmentation framework that integrates DD pulse placement with the routing and scheduling of ion transport. The proposed \emph{shuttling-aware dynamical decoupling} (SADD) pass acts on a resolved QCCD schedule and jointly chooses feasible refocusing pulses and local trajectory changes under transport, site occupancy, and processing zone constraints. At the foundation of this pass, we use a logic- and trajectory-aware quasistatic phase objective to rank candidate schedules. Additionally, this objective can incorporate a known spatial susceptibility profile without requiring noisy circuit simulation. Within our discrete model, accepted edits preserve the logical operations and their timing while only modifying ion trajectories locally.

We implement SADD as a greedy sequence of bounded local optimizations and evaluate it in simulation using one compiled schedule for a set of benchmark circuits. At the tested slow-noise baseline, SADD lowers the finite-correlation dephasing exponent and can lower infidelity when dephasing dominates over control errors. The reduction of the dephasing exponent weakens as temporal or spatial correlations shorten, while modeled control and transport penalties can remove the fidelity gain.
Supplying a spatial susceptibility profile lowers the targeted dephasing exponent further and can improve fidelity while transport remains cheap.

These results establish a controlled proof of concept within the tested linear architecture and phenomenological noise models and show that this schedule-level formulation allows DD to be optimized directly where its pulse timing is realized. Finally, this work turns remaining transport freedom into a design variable whose dephasing benefit can be weighed against the physical and computational cost of using it.

This work first reviews QCCD compilation, dynamical decoupling, and position-dependent dephasing in \cref{sec:background}. We then isolate the unresolved gap between QCCD scheduling and compiler-level DD in \cref{sec:bg_related_work}. Following this, we derive the trajectory-aware phase objective in \cref{sec:motivation} and turn it into a bounded schedule-augmentation pass in \cref{sec:method}. Finally, we present the shared numerical protocol and evidence chain in \cref{sec:results}, before \cref{sec:conclusion} discussing the practical scope, limitations, and remaining hardware questions.

\section{Background}
\label{sec:background}

The central gap arises because QCCD compilation and dynamical decoupling address complementary parts of the same control problem. QCCD compilation determines when and where control can be applied, while DD determines how that control should be timed to suppress correlated noise. To make this connection explicit, we first review the shuttling constraints that shape physical control access, the timing condition underlying DD, and the position-dependent dephasing that makes ion transport relevant. We then turn to existing compiler-level approaches and ask to what extent they account for these effects together.

\subsection{Shuttling compilation for QCCD architectures}
\label{sec:bg_qccd}

Quantum charge-coupled device (QCCD) architectures extend trapped-ion processors beyond individual chains by dividing the device into connected trapping regions and shuttling ions between them~\cite{Kielpinski2002QCCD,Bruzewicz2019TrappedIons}. This motion brings selected ions together for local gates and thereby provides effective all-to-all connectivity. In QCCD processors, trapping zones can be arranged along one-dimensional segments~\cite{Kaushal2020Shuttlingbased,Pino2021HoneywellQCCD}, while junctions can connect several such segments into larger layouts~\cite{Lekitsch2017Blueprint,Ransford2026Helios}. The available transport paths, localized processing zones, and shared resources constrain where ions can move and where and when gates can be applied.
Transport may also introduce motional excitation, which can degrade subsequent two-qubit gates~\cite{Walther2012Controlling,Sutherland2022One}.

Together, these architectural constraints and costs require a dedicated shuttling compilation step that translates an input circuit into a feasible and efficient schedule of transport and gate operations. The underlying placement, routing, and scheduling problem is combinatorial, with even the standard qubit-routing subproblem being NP-hard~\cite{Ito2023Algorithmic}. Shuttling compilation produces a resolved physical schedule that determines where and when additional control remains feasible and thereby directly constrains the implementation of dynamical decoupling, whose timing requirements are discussed next.

\subsection{Dynamical decoupling}
\label{sec:bg_dd}

Consider dephasing caused by fluctuations of the qubit transition frequency. In a frame rotating at the nominal qubit frequency, a residual detuning can be modeled by the pure-dephasing Hamiltonian
\begin{equation}
\label{eq:dephasing_hamiltonian}
H_{\mathrm{deph}}(t)
=
\frac{\hbar}{2}\,\delta\omega(t)\,\sigma_z ,
\end{equation}
where $\delta\omega(t)$ is the residual \emph{dephasing rate}. Magnetic-field fluctuations can limit trapped-ion memory coherence~\cite{Harty2014}, while laser-intensity and phase fluctuations add AC-Stark and optical-phase errors~\cite{Schneider1998Decoherence}. A broader review covers other trapped-ion error mechanisms~\cite{Bruzewicz2019TrappedIons}.

A common way to mitigate slowly varying dephasing is \emph{dynamical decoupling} (DD), which applies additional controls to refocus the accumulated phase~\cite{Viola1999Dynamical,Yang2011Preserving}. An $X$-axis $\pi$ pulse reverses the sign of a longitudinal error because $X\sigma_zX=-\sigma_z$. Using a toggling-frame function $y(t)\in\{+1,-1\}$ to track these sign changes, the phase accumulated during an interval of length $T$ is
\begin{equation}
\label{eq:toggling_frame_phase_temporal}
\phi(T)
=
\int_0^T y(t)\delta\omega(t)\,dt .
\end{equation}
For quasistatic dephasing, a midpoint pulse makes the positive and negative phase contributions cancel exactly, leading to the canonical Hahn-echo mechanism~\cite{Hahn1950Spin}. Cancellation becomes imperfect once the dephasing rate $\delta\omega(t)$ varies between the two halves of the window. Multiple-pulse sequences shape $y(t)$ to realize more sophisticated noise suppression~\cite{Cywinski2008How,Yang2011Preserving,Uhrig2007Keeping}, but still rely on the noise varying slowly relative to the pulse pattern.

Because the noise realization is not known before execution, we treat the detuning statistically. For a classical zero-mean Gaussian process, $\phi(T)$ is also Gaussian, and the coherence function becomes~\cite{Cywinski2008How,Yang2011Preserving}
\begin{equation}
\label{eq:dephasing_exponent_definition}
W(T)=\left\langle e^{-i\phi(T)}\right\rangle=e^{-\chi(T)},
\qquad
\chi(T)=\frac{1}{2}\left\langle\phi(T)^2\right\rangle .
\end{equation}
Substituting the toggling-frame phase from \Cref{eq:toggling_frame_phase_temporal} gives the dephasing exponent
\begin{equation}
\label{eq:dephasing_exponent_temporal}
\chi(T)
=
\frac{1}{2}
\int_0^T dt
\int_0^T dt'\,
y(t)y(t')
C_{\delta\omega}(t,t'),
\end{equation}
where $C_{\delta\omega}(t,t')=\langle\delta\omega(t)\delta\omega(t')\rangle$ is the two-time covariance of the residual detuning.

The modulation can also be characterized in the frequency domain. For the toggling function $y(t)$ over a window $w$, define
\begin{equation}
G_y(\omega)
=
\int_w y(t)e^{i\omega t}\,dt,
\qquad
F_y(\omega)
=
\lvert G_y(\omega)\rvert^2 .
\label{eq:background_filter_response}
\end{equation}
At zero frequency,
\begin{equation}
F_y(0)
=
\left\lvert
\int_w y(t)\,dt
\right\rvert^2 ,
\end{equation}
so quasistatic cancellation is equivalent to suppressing the zero-frequency response~\cite{Cywinski2008How,Yang2011Preserving}.

\Cref{eq:dephasing_exponent_temporal} exposes the two ingredients on which DD relies: the control sign pattern determines how phase contributions combine, while the noise covariance determines how strongly those contributions are correlated. In a QCCD architecture, ion transport adds a spatial dependence because a moving ion samples the noise at different locations along its trajectory. We next establish a physical mechanism for this dependence.

\subsection{Position-dependent dephasing in trapped ions}
\label{sec:bg_position_dephasing}

Static spatial variation of a qubit frequency can be calibrated and compensated, but a time-dependent perturbation acting through a nonlinear frequency response can still produce a position-dependent residual. Consider a hyperfine clock transition with quadratic Zeeman response $\omega_q(B)\simeq\omega_{\mathrm{hf}}+\kappa B^2$. Spatially resolved magnetometry can characterize the static field profile $B_{\mathrm{s}}(r)$~\cite{Ruster2017Entanglementbased}, and the resulting position-dependent frequencies can be compensated through spatially aware phase tracking~\cite{RyanAnderson2022Implementing}.

After this known static contribution is accounted for, a small, approximately spatially common fluctuation $\delta B(t)$ leaves the untracked detuning
\begin{equation}
\begin{aligned}
\delta\omega(r,t)
&=
2\kappa B_{\mathrm{s}}(r)\delta B(t)
+\kappa\delta B(t)^2
\\
&\approx
2\kappa B_{\mathrm{s}}(r)\delta B(t),
\end{aligned}
\label{eq:quadratic_zeeman_residual}
\end{equation}
where the last step assumes \mbox{$\lvert\delta B(t)\rvert\ll\lvert B_{\mathrm{s}}(r)\rvert$}. This mechanism has been proposed as a dominant source of low-frequency memory error in trapped-ion clock-state qubits~\cite{Watkins2026Improving}. This provides a concrete physical mechanism for the position-dependent susceptibility used below. Together with the timing dependence of DD, it shows why the realized schedule matters in two ways: it determines both which controls are available and which susceptibility each moving ion samples. The next section reviews how existing compiler-level DD methods account for these schedule-level effects.

\section{The gap between QCCD and compiler-level DD}
\label{sec:bg_related_work}

We next review existing compiler-level approaches to dynamical decoupling and contrast them with existing QCCD compilation methods. This comparison identifies a gap between optimizing DD pulse placement and resolving the ion transport and control constraints that ultimately determine whether those pulses can be realized.

At the circuit level, one class of DD methods treats idle intervals as insertion windows and decides where a standard pulse sequence is useful. Experiments on superconducting processors show that the preferred sequence and insertion rule depend on why a qubit is idle, on surrounding two-qubit activity, and on the cost of the added pulses~\cite{Niu2022Analyzing,Niu2022Effects}. Application-specific selection can instead be learned from structurally similar decoy circuits, allowing DD to be restricted to qubits that show a benefit~\cite{Das2021ADAPT}. These methods make pulse insertion responsive to circuit context or measured device behavior, but their decisions remain pulse selection and placement within circuit-level idle windows.

More expressive compiler methods alter or coordinate control across the circuit. Existing single-qubit gates can be shifted within circuit slack, optionally together with DD, without lengthening the circuit~\cite{Smith2022TimeStitch}. Calibrated coherent and correlated errors, device connectivity, and layer activity can guide a choice between DD patterns and coherent-error compensation~\cite{Seif2024Suppressing}. Graph-based formulations can embed a minimum number of coordinated refocusing pulses for quasistatic single-qubit dephasing and $ZZ$ idling crosstalk~\cite{Coote2025ResourceEfficient}, while syncopated patterns can suppress both single-qubit decoherence and unwanted two-qubit couplings~\cite{Evert2025Syncopated}. These results establish that useful DD placement can depend on the full circuit and hardware context. Their models nevertheless treat qubits as stationary and do not represent transport or the restriction of driven control to localized processing zones.

QCCD compilers solve the complementary physical problem. They optimize transport-aware placement, routing, and schedule duration~\cite{Murali2020Architecting,Saki2022Muzzle,Upadhyay2022ShuttleEfficient,Schoenberger2025MQTIonShuttler}, while architecture-level studies vary topology and resource assignment~\cite{Ovide2024Scaling}. The resulting schedules determine the ion trajectories and the physical control intervals available around logical operations, but are not optimized for DD.

Recent work at the interface of QCCD scheduling and DD demonstrates that the realized transport schedule can directly affect DD performance. Hardware experiments show that scheduling-induced offsets from nominal pulse times weaken low-frequency noise suppression, which can be partially mitigated by compile-time timing feedback and opportunistic real-time insertion~\cite{Watkins2026Improving}. These results make the realized QCCD schedule part of the DD problem and show that DD can be adapted to the timing produced by QCCD compilation.

Overall, the literature still falls into two largely separate lines of work. Compiler-level DD methods optimize pulse placement without accounting for ion transport, while QCCD compilers optimize transport without a DD objective. Watkins et al.\ already connect these two sides, but their scope is experimental mitigation rather than joint ahead-of-time optimization of pulse placement and ion transport. All together, this leads us to the question asked here: Can DD pulse placement and ion transport be optimized jointly?

\section{Phase objective for QCCD schedules}
\label{sec:motivation}

In order to optimize DD pulse placement and ion transport jointly, we need to define a compile-time objective that ranks candidate schedules by their dephasing suppression. Here, we first identify how motion couples pulse feasibility to phase accumulation and then derive such a schedule-level phase objective via a sequence of stated model assumptions.

\subsection{Pulse--trajectory coupling}

The DD formalism in \cref{sec:bg_dd} separates the available control modulation $y(t)$ from the correlation structure of the noise $\delta\omega(t)$. In a resolved QCCD schedule, however, both are trajectory dependent: Pulse placement is constrained by processing-zone availability as well as compatibility with logical gates and transport operations, so a desired modulation $y(t)$ may not be physically realizable. At the same time, a moving ion samples the detuning $\delta\omega(r(t),t)$ along its compiled trajectory $r(t)$. Consequently, a nominal circuit-level pulse placement need not produce the intended physical decoupling after QCCD compilation. 

\begin{figure}[!t]
    \centering

    \includegraphics[width=\columnwidth]{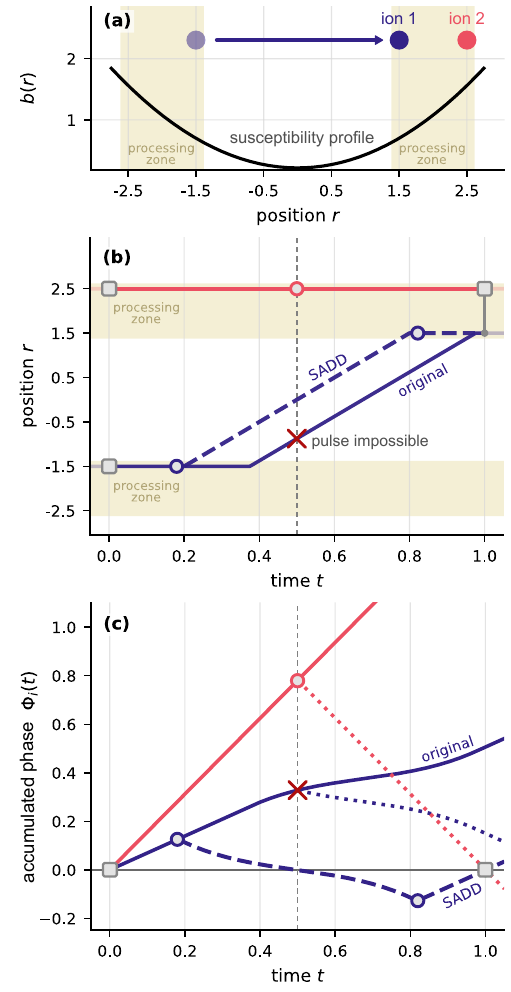}
    \caption{\textbf{Ion trajectories determine decoupling pulse availability and noise exposure.} \textbf{(a)} Ion transport subject to a quadratic susceptibility profile $b(r)$, facilitating shared localized control access between the ions. \textbf{(b)} Ion trajectories with rectangular markers indicating logical gates. A decoupling pulse (round marker) at the window midpoint is optimal for the stationary ion 2 but suboptimal and infeasible for the moving ion 1 due to processing zone access. Rerouting it along the dashed SADD trajectory symmetrizes the susceptibility it samples and admits the two shown pulses. \textbf{(c)} Accumulated quasistatic phase $\Phi_i(t)$ for $\delta\omega(r)=\delta\omega_0 b(r)$ without DD (solid), under a midpoint pulse (dotted), and under the local augmentation (dashed). The augmented trajectory perfectly cancels the accumulated phase of ion 1 by the time of the scheduled two-qubit gate.}
    \label{fig:example}
\end{figure}

\Cref{fig:example} illustrates these coupled effects for an ion shuttled between two processing zones. In this schematic, $\Phi_i(t)$ denotes the signed phase accumulated by ion $i$ along its trajectory up to time $t$. The noise is frozen here, so this is precisely the quantity that \cref{sec:motivation_phase_proxy} generalizes into a compile-time phase proxy.
For simplicity, consider a quasistatic model $\delta\omega(x,t)=\delta\omega_0 b(x)$ with an illustrative phenomenological susceptibility profile $b(x)\propto x^2$, as shown in \cref{fig:example}\textbf{(a)}. Ion 2, which remains in a processing zone throughout the schedule, admits a midpoint Hahn-echo pulse that cancels its phase at the end of the idle window, as indicated by the zero crossing in \cref{fig:example}\textbf{(c)}.
For ion 1, the same midpoint pulse is infeasible because the ion is outside a processing zone at that time, as highlighted in \cref{fig:example}\textbf{(b)}. Even if that pulse were available, it would not cancel the phase because the trajectory samples the spatial susceptibility profile asymmetrically, accumulating more phase in the first half than in the second.
The alternative trajectory and pulse placement shown by the dashed path in \cref{fig:example}\textbf{(b)} are both feasible and perfectly phase cancelling in this example: the longer interval near the less sensitive center compensates for the larger phase accumulated elsewhere.

The simplified example isolates the schedule-level coupling. Equal time intervals need not produce equal phase accumulation, and a phase-balancing pulse time need not be physically available. We therefore require an objective that evaluates the realized pulse pattern and trajectory together while remaining cheap enough to use inside schedule optimization.

\subsection{Trajectory-weighted phase proxy}
\label{sec:motivation_phase_proxy}

We obtain the compile-time phase proxy by extending the temporal covariance in \cref{eq:dephasing_exponent_temporal} along ion $i$'s trajectory and then reducing its calibration and evaluation cost. Replacing the temporal covariance by the space-time covariance sampled along the trajectory gives
\begin{equation}
\label{eq:dephasing_exponent_trajectory}
\chi_i(T)
=
\frac{1}{2}
\int_0^T dt
\int_0^T dt'\,
y_i(t)y_i(t')
C_{\delta\omega}\!\left(r_i(t),r_i(t'),t,t'\right).
\end{equation}
where $C_{\delta\omega}(r_i(t),r_i(t'),t,t')=\langle\delta\omega(r_i(t),t)\delta\omega(r_i(t'),t')\rangle$ is the covariance between the detunings sampled at two points on the trajectory.

Evaluating this full covariance for every candidate would defeat its compile-time role. Motivated by the position-dependent coupling in \cref{sec:bg_position_dephasing}, we therefore use the rank-one model
\begin{equation}
\label{eq:factorized_dephasing_model}
\delta\omega(r,t)
=
\delta\omega_0 b(r)\xi(t),
\end{equation}
where $b(r)$ is an RMS-normalized susceptibility profile and $\xi(t)$ is a zero-mean, unit-variance Gaussian process. For the quadratic-Zeeman mechanism in \cref{eq:quadratic_zeeman_residual}, this corresponds to $b(r)=B_{\mathrm{s}}(r)/\bar B_{\mathrm{s}}$, $\xi(t)=\delta B(t)/\sigma_B$, and $\delta\omega_0=2\kappa\bar B_{\mathrm{s}}\sigma_B$, with $\bar B_{\mathrm{s}}=\sqrt{\langle B_{\mathrm{s}}(r)^2\rangle_r}$ and $\sigma_B=\sqrt{\langle\delta B(t)^2\rangle_t}$. The derivation permits a signed $b(r)$, whereas the implementation and evaluations below restrict it to a nonnegative susceptibility magnitude. This scheduling model makes several assumptions compared to a general description of QCCD noise, in particular, the common-mode fluctuation, small-fluctuation linearization, rank-one spatial factorization, and Gaussian statistics.

For compile-time ranking, we further take $\xi(t)$ to be quasistatic over the interval. This motivates the \emph{phase proxy}
\begin{equation}
\label{eq:continuous_phase_proxy}
\Phi_i(T)
=
\delta\omega_0
\int_0^T
y_i(t)b(r_i(t))\,dt .
\end{equation}
The realized phase of \cref{eq:toggling_frame_phase_temporal} is then $\phi_i(T)=\xi\Phi_i(T)$, so $\Phi_i$ is a phase amplitude rather than a realized phase, and \cref{eq:dephasing_exponent_definition} gives $\chi_i(T)=\Phi_i(T)^2/2$. Thus, minimizing $\lvert\Phi_i(T)\rvert$ minimizes the dephasing exponent for one ion over one interval under this model. A circuit-level objective must still determine where scalar phase accumulation can continue through logical gates and where it must be partitioned.

\subsection{Logic-aware phase objective}
\label{sec:logic_critical_windows}

To extend this single-window proxy across a circuit, we partition the schedule according to how each logical operation transforms a longitudinal error. For a unitary $U$ acting on ion $i$, we continue scalar accumulation when
\begin{equation}
U \sigma_z^{(i)} U^\dagger = s_i(U)\sigma_z^{(i)},
\qquad s_i(U)\in\{+1,-1\},
\end{equation}
where the operation multiplies the toggling sign by $s_i(U)$. If the condition does not hold, we close the scalar accumulation immediately before $U$ and start a new window at $U$, the boundaries of which we call \emph{critical points}. This construction partitions the schedule for the purpose of defining a cheap objective such that it does not physically remove the earlier error or propagate the resulting operator through the remainder of the circuit.

The intervals between consecutive critical points define the \emph{critical windows} $\mathcal W_i$ for ion $i$, based on which we define the schedule-level \emph{phase objective}
\begin{equation}
\label{eq:phase_proxy_global_crit_points}
    J_\Phi
    =
    \sum_{i \in Q}
    \sum_{w \in \mathcal{W}_i}
    \Phi_i(w)^2,
\end{equation}
with the per-window phase proxy
\begin{equation}
\label{eq:crit_points_phase_proxy}
\Phi_i(w)
=
\delta\omega_0
\int_w
y_i(t)b(r_i(t))\,dt .
\end{equation}
Here $Q$ is the set of qubits, $y_i(t)$ includes sign changes from refocusing pulses and scalar-preserving logical operations, and $r_i(t)$ is the compiled trajectory of ion $i$.
Thus, $J_\Phi$ is our gate-partitioned scheduling cost, obtained by summing the squared phase proxies of the individual critical windows. The logical gates determine where scalar accumulation is continued or partitioned, but $J_\Phi$ does not model complete logical error propagation.

Applying \cref{eq:background_filter_response} to $q_{i,w}(t)=y_i(t)b(r_i(t))$ gives the zero-frequency identity
\begin{equation}
J_\Phi
=
\delta\omega_0^2
\sum_{i\in Q}\sum_{w\in\mathcal W_i}F_{q_{i,w}}(0).
\end{equation}
Hence, in the factorized quasistatic limit, $J_\Phi$ is the summed zero-frequency response of the critical windows. For one ion and one critical window, the relation $\chi_i(w)=\Phi_i(w)^2/2$ is exact under this model. For a general multiqubit circuit, however, $J_\Phi$ is neither the dephasing exponent of a specified joint coherence nor a state-infidelity functional. It omits error propagation and correlations across mixing gates and critical windows, cross-ion terms under spatially correlated noise, and the state-dependent effect of the resulting error operators. We therefore use $J_\Phi$ only to rank candidate schedules and assess circuit-level benefit independently through end-to-end simulation.

\section{Bounded schedule augmentation}
\label{sec:method}
The phase objective now ranks any resolved schedule, but it does not construct a valid edit that lowers it.
To do so, we implement the proposed \emph{shuttling-aware dynamical decoupling} (SADD) framework as a greedy post-processing pass that applies local pulse-and-trajectory edits to a valid input schedule in order to reduce the phase objective $J_\Phi$.

Two considerations motivate this post-processing approach. First, DD-agnostic shuttling compilation is itself the high-dimensional scheduling and routing problem of \cref{sec:bg_qccd}, and adding pulse selection, placement, and noise-aware routing would further enlarge its search space. Second, a fully joint formulation would make logical gates and refocusing pulses compete for the same control resources, turning short schedules and dense decoupling into conflicting objectives. Augmenting a resolved schedule instead keeps the upstream compilation result authoritative and confines DD to the control resources it leaves unused. The pass is therefore organized around remaining \emph{control opportunities} rather than qubit idleness, and belongs after the first compilation layer that exposes resolved trajectories, operation durations, and control-resource occupation, while local pulse and trajectory edits are still available. Here, we first define the discrete schedule model and the invariants an accepted edit preserves, then describe the local optimizer that proposes candidates.

\subsection{Schedule model and invariants}
\label{sec:method_architecture_abstraction}

To define schedule-preserving edits, we instantiate the pass on the linear segmented architecture reviewed in \cref{sec:bg_qccd}. We model an architecture $\mathcal A$ as a one-dimensional array $\mathcal V$ of discrete sites with processing zones $\mathcal P=\{P_1,\ldots,P_K\}$, where each $P_k\subseteq\mathcal V$ is a contiguous subset of sites. Time is discretized into layers of duration $\Delta t$. At each timestep, each site can host at most one ion.

The considered abstraction supports nearest-neighbor shuttles, adjacent ion exchanges, and gates from
\begin{equation}
\mathcal G =
\left\{
R_x(\theta),
R_y(\theta),
R_z(\theta),
R_{zz}(\theta)
\right\}.
\label{eq:native-gate-set}
\end{equation}
We use the half-angle conventions
\begin{equation}
\begin{aligned}
R_\alpha(\theta)
&=\exp\!\left(-\frac{i\theta}{2}\sigma_\alpha\right),
&&\alpha\in\{x,y,z\},\\
R_{zz}(\theta)
&=\exp\!\left(-\frac{i\theta}{2}\sigma_z\otimes\sigma_z\right).
\end{aligned}
\end{equation}
Here, $R_x$ and $R_y$ are driven single-qubit rotations, $R_z$ is an exact virtual frame update, and $R_{zz}$ is the native entangling operation. Circuit inputs are normalized to this gate set before routing, while other entangling axes are rejected rather than treated as native operations.

The physical operations are subject to several constraints. In particular, an ion may shuttle only to a neighboring site that is unoccupied in the next timestep. Within this abstraction, an adjacent exchange swaps the positions of two ions on neighboring sites. Additionally, driven gates may only be performed when all participating ions are located inside the same processing zone. Finally, each processing zone can execute at most one driven gate at a time, and each ion can participate in at most one physical operation per timestep. Virtual $R_z$ updates preserve their exact logical action but consume no time, processing-zone occupancy, or control pulse.

Operation durations are supplied as integer multiples of $\Delta t$ and determine resource occupation in the scheduler. The abstraction assumes that replacing $\theta$ by $-\theta$ leaves the participating ions, duration, and processing-zone requirements unchanged, including for $R_{zz}$. If a backend does not satisfy this contract, the affected frame must be closed or the transformed schedule recompiled, and unchanged timing no longer follows. Lower-level effects such as electrode waveforms, crystal splitting and merging, sympathetic transport, and motional-mode evolution are not modeled explicitly, such that their timing and feasibility requirements are represented by the shuttle, exchange, and gate operations and are assumed to be resolved by a downstream pulse-compilation layer.

The input is a discretized QCCD execution schedule $\mathcal S$ over $\mathcal A$. It specifies each trajectory $r_i[n]\in\mathcal V$ and every scheduled gate, transport, exchange, and idle operation, including participants, logical parameters, durations, start times, and occupied processing zones. A processing-zone control opportunity $u=(P_k,[n_0,n_1))$ is a contiguous interval of timesteps in which $P_k$ is not used by a pre-existing logical gate. Within such an interval, SADD may insert refocusing pulses and modify local trajectories while retaining the fixed boundary conditions imposed by the surrounding schedule.

The input schedule is assumed to be physically valid within the discrete architecture and to implement the intended logical circuit. SADD keeps the logical gate identities, target parameters, order, and start times fixed, together with the discrete makespan and final ion placement. Logical-frame tracking may replace the physical angle $\theta$ used to realize a logical gate by $-\theta$ under the equal-resource contract above.

The augmented schedule can change the known phase accumulated from calibrated position-dependent frequencies. We assume that this phase is compensated when the final schedule is converted to hardware controls. This deterministic correction is separate from the zero-mean stochastic detuning ranked by $J_\Phi$. After this compensation, the only remaining logical-frame change introduced by SADD is the parity of the inserted $X$ pulses, as described below. The reported noise model is formulated in this compensated rotating frame. If compensation consumes scheduled control or changes an operation duration, its resource use and the makespan must be revalidated.

For phase-proxy replay, the critical-window rule from \cref{sec:logic_critical_windows} classifies $R_z$ and $R_{zz}$ as sign preserving, while $R_x(\theta)$ and $R_y(\theta)$ with $\theta\notin\pi\mathbb Z$ form window boundaries. Odd multiples of $\pi$ about either $x$ or $y$ instead flip the toggling sign within the current window, while even multiples preserve it.

Timestep $n$ denotes the half-open interval $[n\Delta t,(n+1)\Delta t)$. Operations assigned to its leading boundary $n\Delta t$ are resolved before the phase increment for that interval. A logical or refocusing gate first updates $y_i[n]$, while a shuttle or exchange first updates $r_i[n]$ to its post-transport site. The interval then contributes $\delta\omega_0\Delta t\,y_i[n]b(r_i[n])$. The local integer objective and floating-point full-schedule replay use this same ordering. For their outputs to be comparable, the numerical noise replay and required deterministic-phase compensation must use it as well. The discrete form of \cref{eq:crit_points_phase_proxy} is
\begin{equation}
\Phi_i(w)
=
\delta\omega_0
\Delta t
\sum_{n\in w}
y_i[n]b(r_i[n]) ,
\end{equation}
where $n$ indexes schedule timesteps in critical window $w$. In the implementation, $b(r)$ is restricted to nonnegative values and normalized to unit RMS over the architecture sites. Because $\delta\omega_0^2$ is common to all candidate schedules, the implementation omits this factor when evaluating $J_\Phi$.

Gates and transport operations reserve their full declared durations, and phase accumulation remains present in those layers. Gate and position updates occur at the leading boundary of each operation, following the layer-level convention defined above. If a later compilation stage changes the timing, ordering, or trajectories, the compensation and schedule validation must be repeated. With physical validity defined within this abstraction, we next specify how inserted pulses preserve the logical computation.

\subsection{Logical-frame tracking}
Inserted refocusing pulses can leave an ion in a nontrivial logical frame, so preserving the intended computation requires propagating that frame through subsequent gates.
SADD represents DD pulses as additional operations using the same controls as logical single-qubit gates. It inserts only $R_x(\pi)$ pulses and therefore tracks a binary logical frame $f_i[n]$ for each ion and schedule layer, indicating the two possible frames $\{I,X\}$. Each ion's frame is initialized in $I$, and every inserted $R_x(\pi)$ pulse toggles the corresponding bit. For the considered native gate set in \cref{eq:native-gate-set}, conjugation by $X$ gives
\begin{equation}
\begin{aligned}
X R_x(\theta) X &= R_x(\theta), \\
X R_y(\theta) X &= R_y(-\theta), \\
X R_z(\theta) X &= R_z(-\theta),
\end{aligned}
\label{eq:single-qubit-frame-updates}
\end{equation}
and a two-qubit gate on ions $i$ and $j$ is updated as
\begin{equation}
R_{zz}(\theta)
\quad \longmapsto \quad
R_{zz}\!\left((-1)^{f_i+f_j}\theta\right).
\label{eq:rzz-frame-update}
\end{equation}
Here $f_i$ and $f_j$ are the incoming frames immediately before the two-qubit gate.

To state the equivalence precisely, let $k$ follow the quantum-control event order used by schedule replay; genuinely simultaneous events with disjoint supports may be ordered arbitrarily. Let $G_k$ be the $k$th logical event, with $G_k=I$ for an inserted DD pulse, and define $U_{\mathrm{log}}^{(k)}=G_k\cdots G_1$. Let $\widetilde G_k$ and $U_{\mathrm{phys}}^{(k)}=\widetilde G_k\cdots\widetilde G_1$ denote the corresponding ideal physical targets after augmentation. If $F_k=\bigotimes_iX_i^{f_i^{(k)}}$ is the frame after event $k$, then, up to the global phases of the inserted pulses, the invariant is
\begin{equation}
U_{\mathrm{phys}}^{(k)}=e^{i\gamma_k}F_kU_{\mathrm{log}}^{(k)}.
\label{eq:logical_frame_invariant}
\end{equation}
It holds initially with $F_0=I$. For a pre-existing logical gate, the frame is unchanged and choosing $\widetilde G_k=F_{k-1}G_kF_{k-1}^{\dagger}$ gives the parameter updates in \cref{eq:single-qubit-frame-updates,eq:rzz-frame-update}. For an inserted pulse on ion $i$, $\widetilde G_k=R_{x,i}(\pi)=-iX_i$ and $F_k=X_iF_{k-1}$. In either case, \cref{eq:logical_frame_invariant} is preserved, so induction gives $U_{\mathrm{phys}}=e^{i\gamma}F_{\mathrm{term}}U_{\mathrm{log}}$, where $F_{\mathrm{term}}=\bigotimes_iX_i^{f_i^{\mathrm{term}}}$.

Final Pauli measurements can be interpreted in this frame by classical outcome processing. For the state-fidelity calculation below, the simulated physical output is expressed in the logical frame as $\rho_{\mathrm{noisy}}^{(\mathrm{log})}=F_{\mathrm{term}}^\dagger\rho_{\mathrm{noisy}}^{(\mathrm{phys})}F_{\mathrm{term}}$ before comparison with the ideal state. If an application requires a physical quantum state in the original laboratory frame rather than frame-adjusted measurements, a nontrivial terminal frame requires compatible downstream control or an explicit compensating operation and is not generally overhead-free. With this conditional logical equivalence defined, the remaining task is to select and optimize bounded schedule neighborhoods.

\subsection{Local optimization}

SADD first identifies the critical windows and scans the input schedule for processing-zone control opportunities. It then processes the opportunities chronologically. For each opportunity $u$, a pre-selection check forms a bounded set $Q_u$ of participating ions that can reach the processing zone between their fixed schedule obligations if routing conflicts are temporarily ignored. The local solver subsequently enforces congestion, collision, and processing-zone resource constraints. When more ions are eligible than the configured bound permits, the deterministic priority rule in \cref{appendix:sat_details} selects the participating subset.

The commitments $C_u$ collect everything that the local problem must preserve such as the participating ions' boundary positions, all pre-existing logical gates, and any transport or exchange operation coupling a participating ion to a non-participating ion. These commitments restrict the decision variables to pulse insertions and trajectory changes inside the opportunity. Ion trajectories are represented by per-ion, per-layer site occupations, and candidate $R_x(\pi)$ pulses by binary variables. The local objective is the contribution of the affected critical windows to $J_\Phi$. We solve the resulting discrete problem with OR-Tools CP-SAT~\cite{cpsatlp}, while the encoding and reported solver settings are given in \cref{appendix:sat_details}.

After decoding, each candidate is replayed against the full schedule and committed only if it satisfies every discrete hardware and schedule constraint and strictly decreases the floating-point value of $J_\Phi$. The pass is therefore greedy over control opportunities and is not globally optimal over all possible trajectory and pulse modifications. Each local instance is limited by a maximum opportunity duration $L_{\max}$, a maximum number of participating ions $N_{\max}$, and a solver time limit $\tau$. For the fixed architecture evaluated here, these limits bound the ion--time part and solver budget of each subproblem; on a larger architecture, the site-variable count may also grow unless the represented site region is bounded. \Cref{appendix:complexity_parallelization} gives the corresponding cost analysis.

Under the valid-input assumption, the equal-resource gate-rewrite contract, the shared replay convention, recomputed deterministic-phase compensation, and compatible handling of the terminal $X$ frame, every committed edit has four guarantees within the discrete abstraction. Fixed logical commitments and boundary positions preserve the logical gate schedule, makespan, and final ion placement. Full-schedule replay verifies the site, transport, ion, and processing-zone constraints. The invariant in \cref{eq:logical_frame_invariant} preserves the ideal computation up to $F_{\mathrm{term}}$, and the floating-point acceptance check ensures a strict decrease of $J_\Phi$. These statements do not establish waveform-level validity. Subject to these interfaces, the bounded chronological pass is a timing-preserving local heuristic. The empirical question is whether this guaranteed objective reduction remains predictive beyond the quasistatic optimization model and ultimately translates into higher logical-frame fidelity after the added control and transport costs are included.


\section{Results}
\label{sec:results}

The optimizer of \cref{sec:method} ranks schedules by the quasistatic phase objective $J_\Phi$ and is blind to the error cost of the additionally inserted controls. Its benefit must therefore be established end-to-end rather than on the objective it optimizes. In the following evaluation, we first ask how effective SADD is and in which noise regimes it is beneficial. Two ablations then audit the method's design: holding ions on their input trajectories isolates what bounded rerouting contributes, and withholding the susceptibility profile isolates what profile information contributes, each weighed against the incurred transport overhead. Finally, we ask how far the quasistatic assumption behind the objective extends and audit the computational cost of the pass.

\subsection{Evaluation Setup}
\label{sec:res_setup}

\subsubsection{Benchmark Instances}
All experiments start from compiled QCCD schedules produced by a DD-agnostic, A*-based shuttling compiler for a linear QCCD architecture. The benchmark corpus uses a fixed architecture with nine sites and two processing zones with two sites each and includes varying circuit structures and sizes, specifically, trotterized Ising evolution, QFT, exact QPE, GHZ, and seeded random circuits at $n\in\{4,5,6,8\}$ qubits each, giving 20 schedules with equal family representation. The circuits are obtained from MQT Bench~\cite{MQTBench}. Schedules use $\Delta t=100\,\mu\mathrm{s}$, with shuttles and driven single-qubit gates occupying $1\Delta t$, adjacent exchanges occupying $3\Delta t$, two-qubit gates occupying $2\Delta t$, and virtual $R_z$ gates consuming no time. SADD bounds each local problem to $L_{\max}=16$ timesteps and $N_{\max}=5$ participating ions.
All schedule generation and SADD augmentation were implemented in MQT IonShuttler~\cite{Schoenberger2025MQTIonShuttler,IonShuttler}, and all noisy circuit simulations were performed with MQT YAQS~\cite{Sander2025Quantum, YAQS}. Both packages are available as part of the Munich Quantum Toolkit~\cite{Wille2024MQT}. All results and timings were generated on a laptop with an Intel Ultra 7 258V CPU.

\subsubsection{Compared Methods}

The proposed method, in the following referred to as \emph{Full SADD}, is compared against four reference methods: \emph{No DD} is the unmodified input schedule and supplies the common non-suppressed reference. As a realizable naive comparator, we consider the \emph{Nearest Hahn} scheme, which attempts to place one $R_x(\pi)$ pulse at the ideal midpoint of every eligible gate-idle window, projecting each of those pulses onto the nearest time at which the compiled trajectory already permits it, dropping pulses when no such time exists (e.g. the ion does not have access to an unoccupied processing zone anywhere in the window). As a further baseline we consider the \emph{Idealized Hahn} scheme, which uses the same midpoint placement scheme while entirely ignoring processing-zone access and ion availability constraints. This idealized counterfactual thus measures what midpoint refocusing would achieve with ubiquitous local control. In parts of the evaluation we additionally consider a modified version of the proposed method that optimizes $J_\Phi$ with only pulse placements on the existing trajectory, without rerouting, which we refer to as \emph{Pulse-only SADD}. The inserted decoupling pulses across all four DD methods carry the same single-qubit gate error detailed below.

\newpage

\subsubsection{Noise models}
The noise falls into two primary categories:
\paragraph{Dephasing} Noisy replays draw a stationary, zero-mean Gaussian Ornstein--Uhlenbeck detuning field with covariance
\begin{equation}
\begin{aligned}
\label{eq:noise_model_covariance}
C_{\delta\omega}(r,r',n,n')
&=
(\delta\omega_0)^2 b(r)b(r')
\\
&\quad\times
e^{-|r-r'|/\ell_c}
e^{-|n-n'|\Delta t/\tau_c}.
\end{aligned}
\end{equation}
Here $\delta\omega_0$ is the dephasing amplitude, $\tau_c$ the temporal correlation time, and $\ell_c$ the spatial correlation length in site units. For each stochastic realization, we jointly sample the detuning field over all sites and layers, which each ion samples along its trajectory. Unless stated otherwise, we take a uniform susceptibility profile $b(r)=1$, $\tau_c=1\,\mathrm{s}$, and $\ell_c\rightarrow\infty$.

\paragraph{Control error} Every driven single- or two-qubit rotation receives an independent, zero-mean fractional pulse-area error of standard deviation $0.001s_{\mathrm{ctrl}}$, while axis error is represented by independent components perpendicular to the intended generator with component width $0.05^\circ s_{\mathrm{ctrl}}$ before normalization. Thus $s_{\mathrm{ctrl}}=1$ denotes $0.1\%$ pulse-area error and $0.05^\circ$ axis tilt. $R_z$ gates are considered virtual and thus remain exact. To relate $s_{\mathrm{ctrl}}$ to a per-pulse error scale, the leading-order mean infidelity of one added $R_x(\pi)$ pulse gives the model-equivalent scale
\begin{equation}
\begin{aligned}
r_{\pi}^{\mathrm{ctrl}}
&\simeq
\frac{\pi^2}{6}(0.001s_{\mathrm{ctrl}})^2
+\frac{4}{3}(0.05^\circ s_{\mathrm{ctrl}})^2
\\
&=2.66\times10^{-6}s_{\mathrm{ctrl}}^2 .
\end{aligned}
\label{eq:control_equivalent_infidelity}
\end{equation}

\begin{figure}[t!]
    \centering
    \includegraphics[width=0.98\columnwidth]{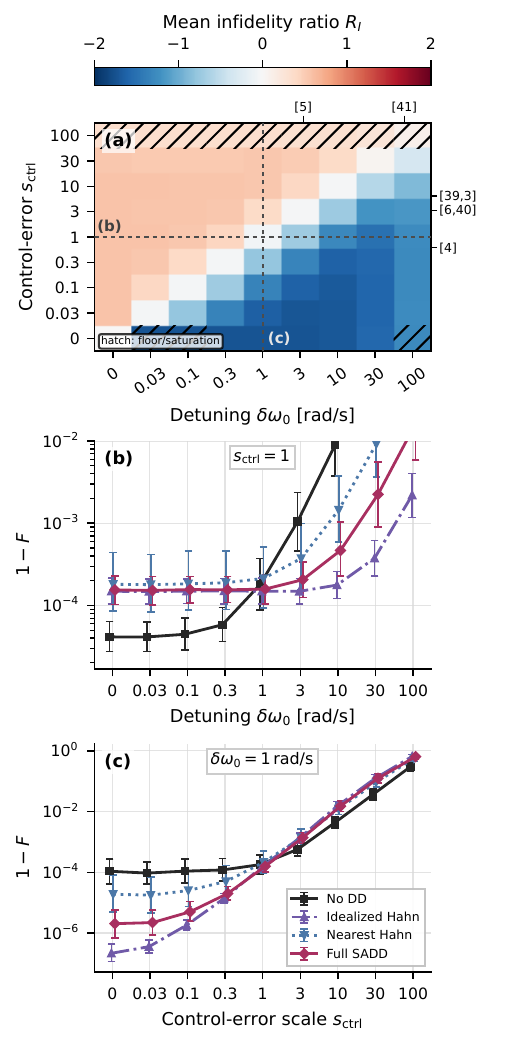}
    \caption{\textbf{SADD improves fidelity in the dephasing-dominated regime and outperforms Nearest Hahn.} \textbf{(a)} Mean Full-SADD/no-DD infidelity log ratio $R_I$ over the detuning scale $\delta\omega_0$ and the pulse-error scale $s_{\mathrm{ctrl}}$, with blue favoring SADD. The two domains meet along the diagonal. Bracketed ticks locate published device measurements on each axis~\cite{Harty2014,Pino2021HoneywellQCCD,Ransford2026Helios,Hilder2022FaultTolerant,QuantinuumH2DataSheet,Ruster2016longlived,Brandl2016Cryogenic}, and hatched settings contain circuits at the resolvable infidelity floor or at saturation and should not be read quantitatively. The dashed lines mark the two cross-sections below. \textbf{(b)} Geometric-mean absolute infidelity along the $s_{\mathrm{ctrl}}=1$ row, placing Full SADD between the realizable and the idealized Hahn baseline. \textbf{(c)} The same quantity along the $\delta\omega_0=1\,\mathrm{rad\,s^{-1}}$ column, sweeping the pulse-error scale instead.}
    \label{fig:operating_regime}
\end{figure}

\subsubsection{Fidelity Metrics}

To investigate the trade-off between dephasing suppression and control error, final state infidelity is quantified as
\begin{equation}
I=1-F,
\qquad
F=\langle\psi_{\mathrm{ideal}}\rvert
\rho_{\mathrm{noisy}}
\lvert\psi_{\mathrm{ideal}}\rangle ,
\end{equation}
where $\lvert\psi_{\mathrm{ideal}}\rangle$ is the noiseless final state of the DD-free logical circuit and $\rho_{\mathrm{noisy}}$ is the ensemble-averaged noisy final state after transformation into the tracked logical terminal frame.

\begin{figure*}[t]
    \centering
    \includegraphics[width=\textwidth]{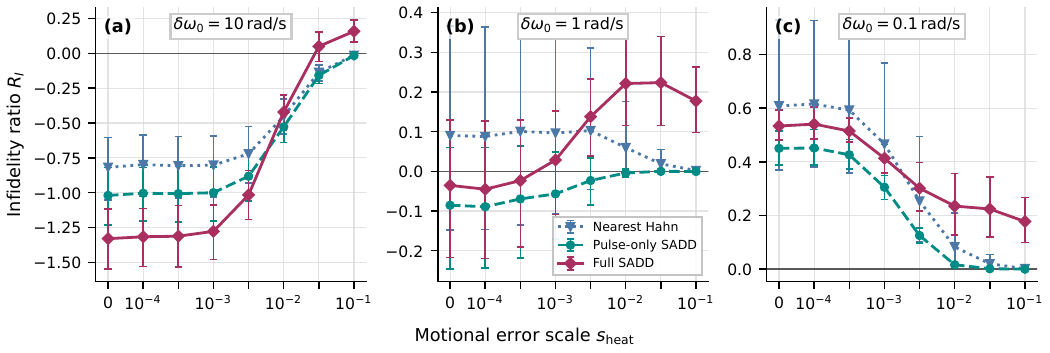}
    \caption{\textbf{Ion rerouting is beneficial while transport is cheap.} State-infidelity log ratios $R_I$ against a common no-DD reference over the motional-error scale $s_{\mathrm{heat}}$ of \cref{eq:heating_error}, under a uniform susceptibility profile at $s_{\mathrm{ctrl}}=1$ and three detunings: \textbf{(a)} $\delta\omega_0=10$, \textbf{(b)} $1$, and \textbf{(c)} $0.1\,\mathrm{rad\,s^{-1}}$. Negative values improve on No DD. Each panel is scaled to its own data, since the benefit spans decades across the three settings.}
    \label{fig:rerouting_benefit}
\end{figure*}

For any positive quantity $X$ we report paired log ratios
\begin{equation}
R_X(A/B)
=
\log_{10}\frac{X_A}{X_B},
\label{eq:evaluation_log_ratio}
\end{equation}
so that $R_X<0$ favors method $A$. For brevity we write $R_J$ for the log ratio of the phase objective $J_\Phi$ and $R_\chi$ for that of the summed dephasing exponent $\chi_\Sigma$ defined in \cref{eq:summed_dephasing_exponent}. Every noise setting is sampled with 64 stochastic realizations per circuit, matched across schedule variants, and the reported center is the unweighted mean of the per-circuit log ratios. Wherever a figure shows an interval or a shaded band, it is the $95\%$ interval obtained by resampling the benchmark circuits while preserving the method pairing.

\subsection{Fidelity benefit and operating regime}
\label{sec:res_regime}

Adding refocusing pulses suppresses dephasing but exposes the computation to additional control error. We therefore first identify the relative noise regime in which SADD is beneficial and then compare its performance to other DD methods.

\Cref{fig:operating_regime}\textbf{(a)} reveals that the noise plane is split into two domains along the diagonal where $\delta\omega_0$ and $s_{\mathrm{ctrl}}$ are the same magnitude. In the dephasing-dominated regime, i.e. where $\delta\omega_0$ is stronger than $s_{\mathrm{ctrl}}$, Full SADD improves fidelity over No DD, achieving up to $50$ times lower infidelity in the limit of error-free pulses. For the regime where $\delta\omega_0$ is weaker than $s_{\mathrm{ctrl}}$, i.e., where control error dominates, it is detrimental throughout, resulting in up to four times larger infidelity. SADD's fidelity benefit thus depends on the relationship between control error and dephasing scales, which matches expectation around inserting imperfect refocusing pulses.

To put these scales in perspective, both axes carry reference ticks derived from the cited measurements. The pulse-error ticks are conservative model-equivalent upper bounds obtained from single-qubit randomized-benchmarking infidelities through \cref{eq:control_equivalent_infidelity}. The detuning ticks are Gaussian-equivalent rms detunings converted from reported Ramsey free-induction decay times using $C(T)=\exp[-(\delta\omega_0T)^2/2]$~\cite{Ruster2016longlived,Brandl2016Cryogenic}. No cited source supplies both coordinates for a single device, so the ticks do not place particular hardware on the map. When considered separately, however, they show that the beneficial domain overlaps realistic scales on both axes.

The preceding investigation outlined a noise-domain boundary between the beneficial and harmful regimes of SADD. We now investigate how the method compares to alternative DD schemes. \Cref{fig:operating_regime}\textbf{(b)} compares the different methods along a cross-section of panel \textbf{(a)}, reporting the respective absolute infidelities. Below $\delta\omega_0=1\,\mathrm{rad\,s^{-1}}$, where this cross-section meets the diagonal domain boundary, every DD method is worse than No DD, and above it all are beneficial, indicating that the discussed domain split applies to all investigated DD methods. Full SADD lies between the two Hahn constructions, with infidelity lower than Nearest Hahn by factors of about $1.8$--$4$ in the beneficial domain, but higher than Idealized Hahn by factors of $1.4$--$6$ over the same range. This shows that SADD's objective-aware placement improves substantially on opportunistic midpoint placement, yet remains short of the ubiquitous-control counterfactual.

We next audit SADD's ability to reroute ions during pulse scheduling and to account for a spatially varying dephasing landscape.

\begin{figure*}[t]
    \centering
    \includegraphics[width=\textwidth]{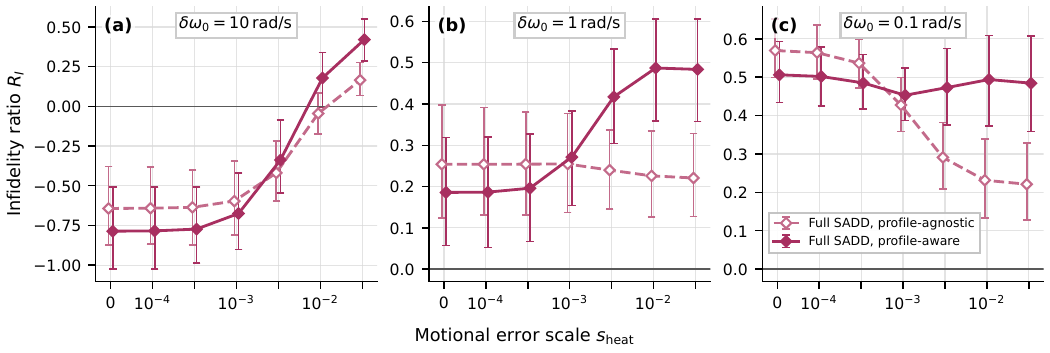}
    \caption{\textbf{Profile-aware compilation is beneficial while transport is cheap.} Profile-aware against profile-agnostic Full SADD under the heterogeneous susceptibility profile of \cref{eq:quadratic_susceptibility_profile}, otherwise as in \cref{fig:rerouting_benefit}: \textbf{(a)} $\delta\omega_0=10$, \textbf{(b)} $1$, and \textbf{(c)} $0.1\,\mathrm{rad\,s^{-1}}$. All three detunings share one compiled schedule set.}
    \label{fig:profile_awareness}
\end{figure*}

\subsection{Benefit and cost of ion rerouting}
\label{sec:res_trajectory}

Within each control opportunity, SADD not only places decoupling pulses but may also reroute ions locally in order to enable better pulse placements. We now investigate when this additional freedom actually helps, and identify the feature's limitations. We therefore compare Full SADD to Pulse-only SADD, which optimizes the same phase objective $J_\Phi$ using the same solver but fixing every ion to its input trajectory, as well as to the Nearest Hahn baseline.

\Cref{tab:method_resources} collects the resulting schedule changes. On fixed trajectories, the objective-aware selection of Pulse-only SADD already places more pulses and reaches a lower phase objective than the opportunistic midpoint choice of Nearest Hahn. Admitting bounded rerouting extends both further. About a third of the pulses placed by Full SADD have no counterpart on the input trajectories because the ion was either outside a processing zone or mid-transport when the pulse was due. Rerouting thus creates control opportunities that the compiled schedule does not offer at all. This comes at the cost of an overall increase in the number of scheduled transport actions.

\begin{table}[t]
\caption{\label{tab:method_resources}
\textbf{Pulse insertions and transport changes by method.} 
Corpus totals over the twenty schedules. Idealized Hahn fills each of the 1122 eligible
gate-idle windows with one pulse; $N_{\mathrm{pulses}}$ and the placement split are
given against that. The difference in transport count $\Delta N_{\mathrm{tr}}$ and the objective ratio $R_J$ are taken against No DD under
a uniform susceptibility profile, which excludes the profile-aware phase objective from comparison.
}
\centering
\footnotesize
\setlength{\tabcolsep}{2pt}
\begin{tabular}{lrcrr}
\toprule
Method & $N_{\mathrm{pulses}}$ & \shortstack{exact/shifted/\\skipped} & $\Delta N_{\mathrm{tr}}$ & $R_J$ \\
\midrule
Idealized Hahn    & $1122$    & $100.0/0.0/0.0\%$    & $0\%$     & $-2.52$ \\
Nearest Hahn      & $-32.7\%$ & $38.6/28.7/32.7\%$   & $0\%$     & $-0.64$ \\
Pulse-only SADD   & $-12.2\%$ & ---                  & $0\%$     & $-0.79$ \\
Full SADD         & $+13.1\%$ & ---                  & $+18.8\%$ & $-1.29$ \\
Full SADD (prof.-aw.) & $-2.7\%$ & ---               & $+62.5\%$ & --- \\
\bottomrule
\end{tabular}
\end{table}

To test whether the additional transport introduced by SADD can offset its dephasing benefit through this error channel, we add a phenomenological stress test. Adapting the count-based heating model from QCCDSim~\cite{Murali2020Architecting}, we introduce a counter $c_i$ for each ion that increments for every shuttle or physical swap in which the ion participates. For a two-qubit gate on ions $i$ and $j$, the model then adds 
\begin{equation}
\delta\theta_{\mathrm{heat}}
\sim
\mathcal N\!\left(
0,
\left[s_{\mathrm{heat}}\frac{c_i+c_j}{2}\right]^2
\right)
\label{eq:heating_error}
\end{equation}
to the gate-angle error, with an error scale $s_{\mathrm{heat}}$ in radians per counted transport participation. In the same small-rotation approximation, its mean added two-qubit-gate infidelity is
\begin{equation}
\overline r_{\mathrm{2q}}^{\mathrm{heat}}
\simeq
\frac{s_{\mathrm{heat}}^2\langle\bar c_g^2\rangle}{5},
\qquad
\bar c_g=\frac{c_i+c_j}{2}.
\label{eq:heating_equivalent_infidelity}
\end{equation}
After the gate, the counters reset. This count-based surrogate does not model persistent motional excitation or explicit cooling, and $s_{\mathrm{heat}}$ should therefore be interpreted as a phenomenological error scale rather than a calibrated heating rate.

\Cref{fig:rerouting_benefit}\textbf{(a)} sweeps this penalty at a representative dephasing-dominated operating point. While transport is cheap, the fidelities follow the phase objective, with Full SADD ahead of Pulse-only SADD and both ahead of Nearest Hahn. The placements that rerouting unlocks therefore do translate into fidelity. As the penalty grows, all three benefits diminish. The fixed-trajectory methods approach No DD, while Full SADD's additional transport causes it to lose its lead and eventually become detrimental. Rerouting thus exchanges a control-availability constraint for a transport budget, and that exchange only pays off while the error charged per transport action stays small against the dephasing the extra pulses remove.

At the domain boundary shown in panel \textbf{(b)}, Pulse-only SADD provides the largest mean fidelity gain, improving over No DD on average while Nearest Hahn does not. This shows that objective-aware pulse selection can retain a fidelity benefit near the domain boundary even without rerouting. In the control-error-dominated setting of panel \textbf{(c)}, none of the three DD methods is viable.

\begin{figure}[t]
    \centering
    \includegraphics[width=\columnwidth]{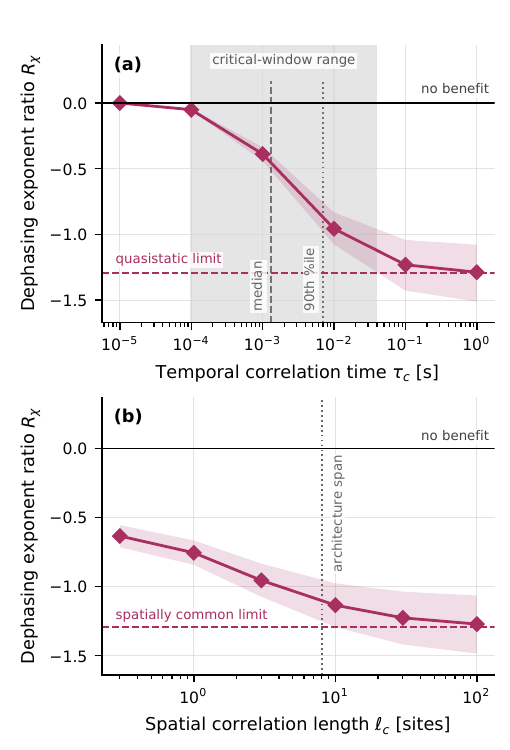}
    \caption{\textbf{The phase objective requires slow noise relative to critical-window length but is robust to short-ranged spatial variation.} Mean dephasing-exponent log ratio $R_\chi$ of Full SADD relative to No DD, shown against \textbf{(a)} the temporal correlation time $\tau_c$ and \textbf{(b)} the spatial correlation length $\ell_c$. Negative values indicate better protection against dephasing. The objective's predictiveness breaks down for correlation times $\tau_c$ that are of the same order as or smaller than the critical-window durations, while it remains robust to correlation lengths even below the site spacing.}
    \label{fig:proxy_applicability}
\end{figure}

\subsection{Value of spatial profile information}
\label{sec:res_profile}

The susceptibility to dephasing may vary spatially across a device, as captured by the profile $b(r)$ in \cref{eq:factorized_dephasing_model}. The preceding evaluations assume a uniform susceptibility profile, $b(r)=1$, but SADD can incorporate a known spatial profile when ranking pulse and trajectory edits. We now compare schedules compiled with this information, referred to as \emph{profile-aware}, against the existing schedules compiled under $b(r)=1$, which we call the \emph{profile-agnostic} baseline. For this comparison we use the qualitatively parabolic profile
\begin{equation}
\label{eq:quadratic_susceptibility_profile}
b(r)\propto(16,8,4,1,0,1,4,8,16),
\end{equation}
motivated by the shape of measured spatial magnetic-field variations~\cite{Ruster2017Entanglementbased}. We normalize it to unit RMS so that its overall noise scale remains comparable to the uniform profile, and replay both variants under this same nonuniform profile.

Awareness of the profile changes which local edits look attractive to the optimizer. Besides inserting pulses, the optimizer can lower the accumulated phase by routing an ion through a less susceptible region. \Cref{tab:method_resources} shows it shifting its effort accordingly, placing fewer pulses than the profile-agnostic variant while spending considerably more transport. We therefore evaluate the resulting trade-off using the same transport-heating stress test as in the preceding section.

In the strong-dephasing regime shown in \cref{fig:profile_awareness}\textbf{(a)}, profile-aware SADD improves over profile-agnostic SADD where transport-induced error is low, showing that the spatial information is beneficial. Once transport becomes expensive, the extra motion cancels that gain and the profile-aware variant degrades more quickly. The crossover, however, occurs only once SADD itself ceases to improve over No DD. None of the tested heating values shows the profile-agnostic variant both outperforming the profile-aware one and still improving over No DD.

The two weaker dephasing settings of \cref{fig:profile_awareness}\textbf{(b)} and \textbf{(c)} lie fully in the detrimental regime. This shift relative to the uniform case occurs because the trajectories favor the center of the architecture, where the ions sample weaker dephasing on average and control error consequently becomes dominant. Profile awareness still lowers the mean infidelity relative to the profile-agnostic schedules while transport is cheap, but not enough for either variant to improve over No DD.

\subsection{Validity range of the phase objective}
\label{sec:res_correlations}

The phase objective $J_\Phi$ is derived for noise that remains static over a critical window and is perfectly correlated across the device. We now ask whether it still selects useful schedules when either assumption is relaxed. To isolate dephasing from control and transport errors, we use the summed dephasing exponent $\chi_\Sigma$ accumulated across all critical windows,
\begin{equation}
\chi_\Sigma=\sum_{i\in Q}\sum_{w\in\mathcal W_i}\chi_i(w),
\label{eq:summed_dephasing_exponent}
\end{equation}
where $\chi_i(w)$ is the dephasing exponent of \cref{eq:dephasing_exponent_definition} for ion $i$ during critical window $w$, evaluated under the finite-correlation noise model of \cref{eq:noise_model_covariance}. The sum reduces to $\chi_\Sigma=J_\Phi/2$ in the quasistatic and perfectly correlated limit, so $\chi_\Sigma$ generalizes the phase objective to finite correlations. Smaller $\chi_\Sigma$ therefore means that the schedule is better protected against dephasing.

The relevant timescale is set by the critical windows themselves, which last on the order of a millisecond in these schedules. \Cref{fig:proxy_applicability}\textbf{(a)} shows that SADD's benefit vanishes once the detuning fluctuates much faster than that, because the phase contributions on opposite sides of the refocusing pulses are no longer sufficiently correlated to cancel. The advantage develops as $\tau_c$ enters the window-duration range and saturates at the quasistatic value beyond it. Thus, ``slow noise'' here concretely means that the detuning must remain correlated over the critical windows used by SADD's phase objective, which in turn depend on the input circuit and its compiled schedule.

The spatial requirement is weaker, as seen in \cref{fig:proxy_applicability}\textbf{(b)}. Full SADD remains favorable even when the noise is correlated over only a small fraction of the site spacing, although its advantage increases toward the fully correlated limit. We attribute this limited sensitivity partly to ions remaining stationary during substantial portions of their critical windows. During these intervals, an ion repeatedly samples the same local fluctuation even when the noise at different sites is uncorrelated. Spatial decorrelation thus matters only while transport carries the ion between sites. For the tested schedules, the proxy therefore requires temporal correlation within the critical windows while remaining robust to spatial decorrelation across the device.

\subsection{Compilation cost}
\label{sec:res_runtime}

The pass runs offline, so its cost matters only through how it grows with the workload. Each control opportunity is one bounded local CP-SAT instance, and because $L_{\max}$ and $N_{\max}$ cap that instance, a larger circuit adds instances rather than enlarging them. The measurements below use a $1\,\mathrm{s}$ solver budget per opportunity.

\begin{figure}[t]
    \centering
    \includegraphics[width=\columnwidth]{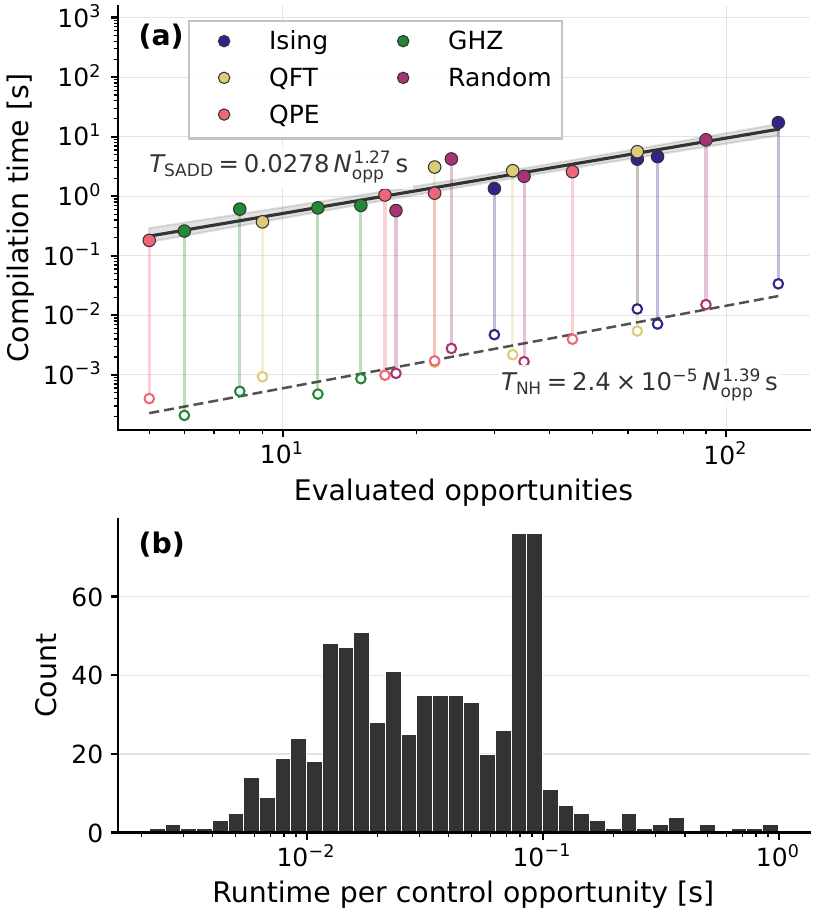}
    \caption{\textbf{Compilation cost is determined by the number of locally bounded control opportunities.} \textbf{(a)} End-to-end per-instance compilation time of Full SADD (filled) and Nearest Hahn (open) against the number of evaluated control opportunities with log--log fits. \textbf{(b)} Distribution of the per-opportunity local-solver runtimes.}
    \label{fig:runtime_scaling}
\end{figure}

\Cref{fig:runtime_scaling}\textbf{(a)} shows the end-to-end compilation time for each schedule as a function of the number of control opportunities evaluated. An empirical fit yields
$T_{\mathrm{SADD}}=0.028\,N_{\mathrm{opp}}^{1.27}\,\mathrm{s}$
with $R^2=0.90$, showing that runtime grows only slightly superlinearly with the number of opportunities over the investigated range. Nearest Hahn is nearly three orders of magnitude faster in the median, but both methods grow at essentially the same rate. The runtime gap therefore represents a constant-factor overhead for SADD relative to the naive comparator rather than a difference in the observed scaling.

As shown in \cref{fig:runtime_scaling}\textbf{(b)}, the individual local solves remain inexpensive, with the median solve taking $32\,\mathrm{ms}$ and $94\%$ finishing within $100\,\mathrm{ms}$. The peak near $10^{-1}\,\mathrm{s}$ reflects the characteristic cost of opportunities that reach the imposed opportunity-duration and participating-ion bounds.

Overall, the compilation cost is governed mainly by the number of opportunities presented by a schedule. Together with the preceding results, this exposes the central tradeoff between SADD's improved dephasing suppression and its additional offline compilation cost.

\section{Discussion}
\label{sec:conclusion}

Dynamical decoupling in QCCD processors is closely coupled to the transport schedule. An ion's route determines when a refocusing pulse can be applied at all, because control is available only in particular zones, and it also determines which dephasing the ion samples on the way. Placing pulses at the circuit level, agnostic of transport, therefore risks timings that are displaced or dropped outright, weakening the intended protection. To address this, we introduced shuttling-aware dynamical decoupling (SADD), a framework for the joint optimization of decoupling-pulse insertion and ion routing, instantiated as a local post-processing pass over an already compiled schedule.

Simulated benchmarks show that, in the slowly varying, dephasing-dominated regime, SADD can substantially improve fidelity over both the original schedule and a simple nearest-feasible Hahn-echo baseline. These gains come with additional transport overhead, particularly when SADD exploits a spatially inhomogeneous dephasing susceptibility, making its effectiveness sensitive to transport error.

The pass is modular with respect to upstream compilation: it runs offline, preserves the gate timings it is given, and does not extend the overall schedule length. Because its edits are local, it can also be applied selectively to chosen intervals of a schedule or around particular processing zones, opening the possibility of targeting only those parts of a computation that are especially exposed to dephasing or that offer sufficient remaining control availability.

The main limitations of this study lie in its hardware and noise abstractions. The discrete operation set represents transport through shuttles and exchanges and does not explicitly capture crystal splitting and merging or waveform-level dynamics. Moreover, the control and transport noise models are not calibrated to a particular device, so the reported crossovers do not represent hardware thresholds. Hardware-facing evaluation therefore requires either a richer transport model with calibrated error parameters or, ideally, direct validation on a QCCD device. Further work could extend SADD with control- and transport-error-aware objective terms, approximate models for residual error propagation beyond critical points, and support for broader DD schemes, including other pulse types and global controls.

Within these bounds, the results establish a proof of concept for shuttling-aware dynamical decoupling in QCCD processors. They show that operating on the physical schedule allows DD pulse placement and ion trajectories to be optimized jointly while accounting for localized control constraints and a spatially varying dephasing landscape.

\begin{acknowledgments}
The project leading to this application/publication has received funding from the European Research Council (ERC) under the European Union’s Horizon 2020 research and innovation program (grant agreement No. 101001318).
This work is also part of the Munich Quantum Valley, which is supported by the Bavarian state government with funds from the Hightech Agenda Bayern Plus, and was further supported by the BMFTR under grant number 13N17298 (SYNQ).
In addition, this work was funded by the Deutsche Forschungsgemeinschaft (DFG, German Research Foundation) – 563436708 \& 563402549.
AI-based tools were used to support language editing and manuscript preparation. All AI-assisted output was critically reviewed and verified by the authors, who take responsibility for the content of the manuscript.
\end{acknowledgments}

\section*{Author Contributions}
L.~S. conceived the work, developed and implemented the SADD method, designed and performed the numerical studies, analyzed the results, and prepared the initial manuscript. A.~S. contributed to the methodology, validation, interpretation of the results, and writing--review and editing. R.~W. supervised the work and contributed to its conceptualization and writing--review and editing. All authors discussed the results and reviewed the manuscript.

\section*{Code Availability}
The code developed specifically for this work is available from the corresponding author upon reasonable request. The open-source software packages used for schedule compilation and noisy simulation, MQT IonShuttler~\cite{IonShuttler} and MQT YAQS~\cite{YAQS}, respectively, are publicly available as part of the Munich Quantum Toolkit~\cite{Wille2024MQT}.

\section*{Data Availability}
The data that support the findings of this study are available from the corresponding author upon reasonable request.

\bibliography{bib}

\newpage 
\setcounter{subsection}{0}
\setcounter{figure}{0}
\renewcommand{\thefigure}{\Alph{section}\arabic{figure}}
\setcounter{table}{0}
\renewcommand{\thetable}{\Alph{section}\arabic{table}}

\appendix

\section{CP-SAT encoding details}
\label{appendix:sat_details}

The local augmentation problem described above is a finite, highly constrained scheduling problem: within a short processing-zone control opportunity, a small set of ions may either retain their original trajectories or be rerouted locally, and additional $X$ refocusing pulses may be inserted if the resulting schedule remains physically admissible. We formulate this subproblem as a constraint-programming satisfiability problem with an integer objective (CP-SAT). This choice is useful here because the relevant degrees of freedom are naturally discrete: ion positions are site occupations, transport corresponds to allowed transitions between adjacent sites, and local controls are binary events at specified processing-zone timesteps.

Consider one processing-zone control opportunity $u=(P,[n_0,n_1))$, where $P\subseteq\mathcal V$ is the set of sites belonging to the processing zone and $[n_0,n_1)$ is a contiguous interval of timesteps during which that zone is not used by an algorithmic gate in the input schedule. The optimizer is applied to a bounded set $Q_u$ of participating ions. These ions are chosen from those that can, ignoring congestion, reach at least one site in $P$ between their fixed schedule obligations. The congestion, collision, and resource constraints are then enforced exactly by the CP-SAT model. Ions not included in $Q_u$ are treated as fixed obstacles following their original trajectories. This keeps each local subproblem small while still allowing the participating ions to exchange positions and expose otherwise unavailable control opportunities.

Timestep $n$ represents the half-open interval $[n\Delta t,(n+1)\Delta t)$. The discrete schedule assigns ion $i$ a single site $r_i[n]$ for phase replay during this interval. Operations scheduled at layer $n$ are resolved before the phase contribution $\delta\omega_0\Delta t\,y_i[n]b(r_i[n])$ is accumulated: logical and refocusing operations update the toggling-frame sign, while a shuttle or swap updates $r_i[n]$ to the post-transport site. Operation durations reserve the corresponding consecutive intervals but do not introduce intermediate transport positions.

After decoding, the proposed local trajectories and inserted pulses are reinserted into the full execution schedule and replayed under the same validity rules as ordinary schedules. Only solutions that pass this full-schedule validation and strictly reduce the phase objective are committed.

\subsection{Decision variables}

For each participating ion $i\in Q_u$, site $s\in\mathcal V$, and relative timestep $m=0,\ldots,L-1$, where $L=n_1-n_0$, we introduce a binary occupation variable
\begin{equation}
    x_{i,s,m}\in\{0,1\},
\end{equation}
which equals one exactly when ion $i$ occupies site $s$ in absolute timestep $n=n_0+m$. The ion trajectory within the local opportunity is therefore represented directly by the set of occupation variables rather than by an explicit list of transport actions. Transport actions are inferred only after the solver has selected a feasible sequence of occupations.

To represent inserted refocusing controls, we introduce a binary variable
\begin{equation}
    c_{i,m}\in\{0,1\},
\end{equation}
where $c_{i,m}=1$ denotes insertion of an $X$ pulse on ion $i$ in the target processing zone at timestep $n_0+m$. The corresponding toggling-frame parity induced by the inserted pulses is represented by another binary variable
\begin{equation}
    f_{i,m}\in\{0,1\}.
\end{equation}
Here $f_{i,m}=0$ and $f_{i,m}=1$ correspond to the two possible signs of the additional local $X$ frame generated inside the opportunity. With the timing convention used for schedule replay, the parity is updated by the inserted pulse at the same timestep,
\begin{equation}
    f_{i,m}=f_{i,m-1}\oplus c_{i,m},
\end{equation}
with $f_{i,-1}=0$ at the beginning of the opportunity. Equivalently, the local contribution of inserted pulses to the dephasing sign is $(-1)^{f_{i,m}}$.

For compactness, the mathematical description below also uses an auxiliary binary movement indicator $a_{i,m}$, which is one when ion $i$ changes site between two consecutive layers. Finally, when the phase objective is embedded into the integer model, auxiliary product variables are introduced to represent products of occupation and frame-parity variables.

\subsection{Hardware and schedule constraints}

The first class of constraints enforces a valid site occupation at every timestep. Each participating ion must occupy exactly one site,
\begin{equation}
    \sum_{s\in\mathcal V} x_{i,s,m}=1
    \qquad\forall i\in Q_u,\; m,
\end{equation}
and each site can host at most one participating ion, after accounting for non-participating ions that remain fixed from the input schedule,
\begin{equation}
    \sum_{i\in Q_u} x_{i,s,m} \leq 1-o_{s,m}^{\mathrm{fix}}
    \qquad\forall s\in\mathcal V,\;m .
\end{equation}
Here $o_{s,m}^{\mathrm{fix}}=1$ if a non-participating ion occupies site $s$ at the corresponding absolute timestep, and zero otherwise. Thus, ions outside the local subproblem remain part of the hardware constraint landscape even though their trajectories are not optimized.

The second class of constraints preserves the interface to the surrounding schedule. At the beginning of the opportunity, each participating ion starts from its position in the input schedule. At the end of the opportunity, it must rejoin the original schedule at the prescribed boundary position. In addition, any pre-existing algorithmic gate involving a participating ion inside the opportunity fixes that ion's position at the corresponding timestep. Transport operations involving a participating ion and a non-participating ion are treated analogously as fixed commitments. These midpoint constraints allow an ion to participate in the optimization even if it has obligations inside the opportunity; the solver may use the free portions before and after such obligations, but it cannot move the gate or change the logical schedule.

Allowed motion is restricted to the local transport graph. For the linear architectures considered here this means that an ion can either remain on its current site or move to a neighboring site in one timestep,
\begin{equation}
    x_{i,s,m}=1,\;x_{i,s',m+1}=1
    \quad\Rightarrow\quad
    |s-s'|\leq 1 .
\end{equation}
Neighboring exchanges are not introduced as a separate decision variable. Instead, if two participating ions occupy adjacent sites in one layer and exchange those sites in the next layer, the decoded trajectory is interpreted as an adjacent exchange. Likewise, a single ion moving into an unoccupied neighboring site is decoded as a shuttle. For operations lasting more than one timestep, the corresponding movement variables are further constrained such that they reserve the participating ions for the full operation duration.

Inserted pulses are constrained by locality and resource availability. A pulse on ion $i$ at timestep $m$ is allowed only if the ion is in the target processing zone and does not move in that layer,
\begin{equation}
    c_{i,m}=1
    \quad\Rightarrow\quad
    \sum_{s\in P} x_{i,s,m}=1,
    \qquad
    a_{i,m}=0 .
\end{equation}
It is also forbidden when the ion is occupied by a pre-existing algorithmic gate or by a fixed transport commitment. Finally, the processing zone itself is treated as a single local control resource for inserted DD pulses,
\begin{equation}
    \sum_{i\in Q_u} c_{i,m}\leq 1
    \qquad\forall m .
\end{equation}
This expresses the assumption that, within the evaluated abstraction, the processing zone can host at most one additional single-qubit refocusing pulse per timestep.

Taken together, these constraints define the set of local schedule augmentations that preserve the external schedule interface, respect site capacity, respect processing-zone control capacity, and remain compatible with the pre-existing logical operations.

\subsection{Objective and local proxy update}

The CP-SAT objective is the local change in the phase objective $J_\Phi$ of \cref{eq:phase_proxy_global_crit_points}. Only critical windows that both involve a participating ion and overlap the local control opportunity can change; all other contributions to $J_\Phi$ are constant for this subproblem.

Let $w=(i,k)$ denote such an affected critical window for ion $i$, and let $\mathcal T_{ik}^{(u)}=\{m\in\{0,\ldots,L-1\}\mid n_0+m\in w\}$ be its timesteps inside opportunity $u$. Its phase proxy is decomposed into pre-opportunity, in-opportunity, and post-opportunity contributions. The pre- and post-opportunity contributions are cached from the current schedule, while the contribution inside the opportunity is determined by the candidate trajectory and inserted-pulse parity selected by the solver. An odd terminal parity reverses the sign of the cached post-opportunity contribution. In discrete time, the candidate phase proxy is
\begin{equation}
\begin{aligned}
    \label{eq:sat_local_phase_update}
    \Phi_{ik}^{(u)}
    &=
    \Phi_{ik}^{\mathrm{pre}} \\
    &+
    \delta\omega_0
    \Delta t
    \sum_{m\in\mathcal T_{ik}^{(u)}}
    \sum_{s\in\mathcal V}
    y_{i,m}^{(0)}
    (-1)^{f_{i,m}}
    b(s)x_{i,s,m} \\
    &+
    (-1)^{f_{i,\mathrm{end}}}
    \Phi_{ik}^{\mathrm{post}} .
\end{aligned}
\end{equation}
Here $f_{i,\mathrm{end}}$ is the parity of the pulses inserted up to the end of opportunity $u$, $y_{i,m}^{(0)}$ is the toggling-frame sign inherited from the current schedule before adding new pulses inside $u$, and $b(s)$ is the RMS-normalized susceptibility profile. The cached phase proxies $\Phi_{ik}^{\mathrm{pre}}$ and $\Phi_{ik}^{\mathrm{post}}$ contain the contributions from the parts of the same critical window before and after the local opportunity, respectively; either contribution is zero when the critical window does not extend beyond the corresponding boundary of $u$.

The primary local objective is then
\begin{equation}
    \min
    \sum_{(i,k)\in\mathcal K_u}
    \left(\Phi_{ik}^{(u)}\right)^2,
    \label{eq:sat_local_objective}
\end{equation}
where $\mathcal K_u$ is the set of affected critical windows. Since $\delta\omega_0$ is common to every candidate, the solver omits its overall square from the objective. CP-SAT represents the remaining real-valued susceptibility profile internally by a fixed integer scale factor. The squared normalized phase proxies in \cref{eq:sat_local_objective} are then ordinary integer quadratic terms implemented through auxiliary integer variables. This scaling only affects the numerical representation; the accepted schedule is re-evaluated using the original floating-point phase objective after decoding. Among solutions with the same phase objective, the solver first prefers fewer inserted pulses and then fewer trajectory changes.

\subsection{Solver configuration and time limits}

The local CP-SAT formulation is used as a bounded post-processing optimizer rather than as a full joint shuttling-and-DD compiler. Long processing-zone idle intervals are split into opportunities of bounded duration, and only a bounded number of participating ions is included in each opportunity. When more ions are eligible than the bound permits, the reported implementation ranks them by their current contribution to $J_\Phi$, followed by their distance from the available control zone and their ion index. Opportunities are then processed in chronological order. In the numerical experiments below we use opportunities of at most 16 timesteps and at most five participating ions, which keeps the local CP-SAT instances small enough to be solved repeatedly across a schedule.

The integer programs are solved with the CP-SAT solver from OR-Tools~\cite{cpsatlp}, using a fixed per-opportunity time limit of 1 s in the reported experiments. A feasible solution is fully replayed and may still be accepted when optimality has not been proven, provided that it is valid and improves the phase objective; opportunities without such a candidate leave the schedule unchanged. The chronological pass is therefore greedy at the level of control opportunities. It is not generally globally optimal over all possible reroutings and pulse placements in the full execution schedule. Instead, it provides a modular way to exploit remaining local control resources while preserving the algorithmic gate schedule, schedule depth in the discrete abstraction, and final ion configuration. This local formulation is also naturally parallelizable at the level of independent candidate opportunities or benchmark schedules, although the results reported here use the outlined chronological iteration scheme.

\section{Circuit-resolved results and schedule changes}
\label{appendix:resource_audit}

\begin{table*}[!t]
\caption{\label{tab:benchmark_resource_audit}
Circuit-resolved benchmark results and compiler audit. Infidelity ratios use 64 matched noise realizations at $\delta\omega_0=10\,\mathrm{rad\,s^{-1}}$ and $s_{\mathrm{ctrl}}=1$; the dephasing-exponent ratios are analytic. In ratio subscripts, $F$, $0$, and $P$ denote full SADD, no DD, and pulse-only SADD; ratios below one favor the numerator method. $N_{\mathrm{tr}}$ is the transport-action count; $N_{\mathrm{pulses}}$ is the number of inserted pulses, with pulses requiring a changed trajectory $N_{\mathrm{nr}}$ in parentheses; $M_{\mathrm{eval}}/M_{\mathrm{acc}}$ counts evaluated/accepted control opportunities; and $t_{\mathrm{SADD}}$ is end-to-end compilation time. Family summary rows report arithmetic means for ratios and totals for counts and time.}
\centering
\scriptsize
\setlength{\tabcolsep}{2.7pt}
\renewcommand{\arraystretch}{1.10}
\begin{tabular*}{\textwidth}{@{\extracolsep{\fill}}llcccccccc@{}}
\toprule
& & \multicolumn{2}{c}{Full SADD/No DD}
& \multicolumn{2}{c}{Full SADD/Pulse-only SADD}
& \multicolumn{4}{c}{Resource audit} \\
\cmidrule(lr){3-4}\cmidrule(lr){5-6}\cmidrule(l){7-10}
Family & $n$
    & $I_F/I_0$ & $\chi_{\Sigma,F}/\chi_{\Sigma,0}$
    & $I_F/I_P$ & $\chi_{\Sigma,F}/\chi_{\Sigma,P}$
& $N_{\mathrm{tr}}^{\mathrm{NoDD}}/N_{\mathrm{tr}}^{\mathrm{Full}}$
& $N_{\mathrm{pulses}}\,(N_{\mathrm{nr}})$
& $M_{\mathrm{eval}}/M_{\mathrm{acc}}$
& $t_{\mathrm{SADD}}$ (s) \\
\midrule
\multirow{5}{*}{Ising}
 & 4              & 0.012 & 0.216 & 1.035 & 0.952 & $48/48$   & $52\ (1)$   & $30/29$  & 1.25 \\
 & 5              & 0.013 & 0.080 & 0.203 & 0.439 & $167/169$ & $103\ (38)$ & $63/59$  & 7.59 \\
 & 6              & 0.005 & 0.019 & 0.379 & 0.090 & $279/275$ & $105\ (29)$ & $70/62$  & 16.01 \\
 & 8              & 0.013 & 0.099 & 0.391 & 0.536 & $628/540$ & $181\ (62)$ & $131/108$ & 43.71 \\
 & \textit{mean/total} & 0.010 & 0.103 & 0.502 & 0.504 & $1122/1032$ & $441\ (130)$ & $294/258$ & 68.56 \\
\midrule
\multirow{5}{*}{QFT}
 & 4              & 0.113 & 0.005 & 1.001 & 0.473 & $14/18$   & $16\ (4)$   & $9/8$    & 1.37 \\
 & 5              & 0.014 & 0.014 & 0.194 & 0.161 & $27/77$   & $54\ (18)$  & $22/21$  & 6.22 \\
 & 6              & 0.015 & 0.015 & 0.105 & 0.077 & $64/162$  & $81\ (41)$  & $33/31$  & 11.74 \\
 & 8              & 0.110 & 0.091 & 0.537 & 0.470 & $163/203$ & $139\ (31)$ & $63/56$  & 26.87 \\
 & \textit{mean/total} & 0.063 & 0.031 & 0.459 & 0.295 & $268/460$ & $290\ (94)$ & $127/116$ & 46.19 \\
\midrule
\multirow{5}{*}{Random}
 & 4              & 0.322 & 0.301 & 0.833 & 0.672 & $28/34$   & $26\ (14)$  & $18/16$  & 1.46 \\
 & 5              & 0.062 & 0.030 & 0.844 & 0.508 & $26/92$   & $55\ (17)$  & $24/20$  & 7.17 \\
 & 6              & 0.105 & 0.052 & 0.416 & 0.217 & $71/135$  & $73\ (29)$  & $35/30$  & 11.98 \\
 & 8              & 0.193 & 0.121 & 0.514 & 0.414 & $163/211$ & $168\ (35)$ & $90/83$  & 27.26 \\
 & \textit{mean/total} & 0.171 & 0.126 & 0.652 & 0.453 & $288/472$ & $322\ (95)$ & $167/149$ & 47.88 \\
\midrule
\multirow{5}{*}{QPE}
 & 4              & 0.273 & 0.062 & 1.669 & 0.106 & $11/13$   & $8\ (5)$    & $5/4$    & 0.45 \\
 & 5              & 0.040 & 0.006 & 0.216 & 0.033 & $33/55$   & $28\ (10)$  & $17/14$  & 2.07 \\
 & 6              & 0.054 & 0.101 & 1.571 & 0.939 & $45/49$   & $31\ (8)$   & $22/15$  & 3.62 \\
 & 8              & 0.142 & 0.069 & 0.439 & 0.376 & $163/181$ & $79\ (19)$  & $45/37$  & 17.51 \\
 & \textit{mean/total} & 0.127 & 0.059 & 0.974 & 0.363 & $252/298$ & $146\ (42)$ & $89/70$ & 23.66 \\
\midrule
\multirow{5}{*}{GHZ}
 & 4              & 0.098 & 0.052 & 0.630 & 0.283 & $8/10$  & $12\ (3)$ & $6/6$   & 0.46 \\
 & 5              & 0.044 & 0.035 & 1.170 & 1.088 & $9/21$  & $18\ (3)$ & $8/8$   & 1.38 \\
 & 6              & 0.052 & 0.085 & 0.333 & 0.271 & $19/37$ & $22\ (6)$ & $12/10$ & 2.16 \\
 & 8              & 0.107 & 0.283 & 0.524 & 0.588 & $37/49$ & $18\ (5)$ & $15/12$ & 2.65 \\
 & \textit{mean/total} & 0.075 & 0.114 & 0.664 & 0.558 & $73/117$ & $70\ (17)$ & $41/36$ & 6.65 \\
\midrule
\multicolumn{2}{l}{\textbf{Total}}
 & --- & --- & --- & ---
 & $2003/2379$ & $1269\ (378)$ & $718/629$ & $192.93$ \\
\bottomrule
\end{tabular*}
\end{table*}

\begin{figure}[!htbp]
    \centering
    \includegraphics[width=\columnwidth]{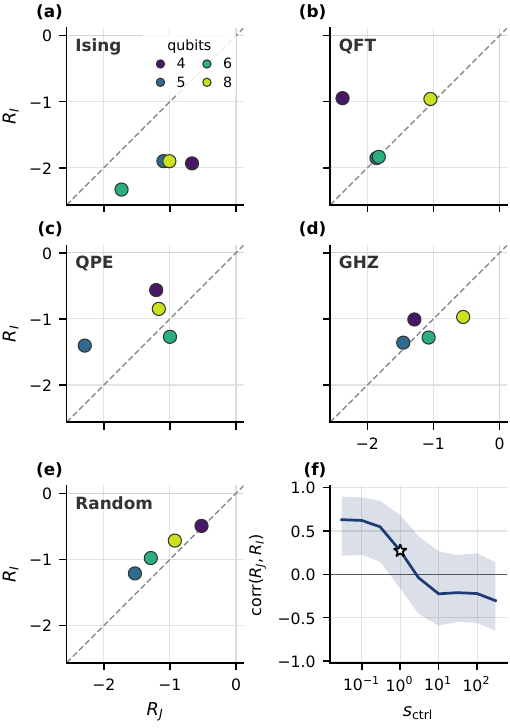}
    \caption{\textbf{The phase objective predicts the direction of the fidelity gain on every circuit, but not its size.} \textbf{(a)}--\textbf{(e)} Per-schedule phase-objective log ratio $R_J$ against state-infidelity log ratio $R_I$, both taken against No DD at $\delta\omega_0=10\,\mathrm{rad\,s^{-1}}$ and $s_{\mathrm{ctrl}}=1$, separated by circuit family. Color gives qubit count, and each panel holds four circuits. Both axes share one range, so the dashed line marks $R_I=R_J$, and schedules below it gain more fidelity than their objective reduction alone would suggest. Two markers overlap in \textbf{(b)}. \textbf{(f)} Pearson correlation between $R_J$ and $R_I$ over all twenty schedules against the pulse-error scale at the same detuning. The star marks the operating point of the other panels, and the band is a $95\%$ circuit bootstrap.}
    \label{fig:objective_fidelity_by_family}
\end{figure}

\Cref{tab:benchmark_resource_audit} gives the results for each benchmark
circuit and summarizes how Full SADD changes the physical schedule, while
the $\chi$ ratios are obtained from the analytic low-frequency calculation.
In all four ratio columns, values below one favor Full SADD.

The remaining columns compare the transport counts before and after Full
SADD and report the number of inserted pulses $N_{\mathrm{pulses}}$, the number
of evaluated and accepted control opportunities
$M_{\mathrm{eval}}/M_{\mathrm{acc}}$, and the Full-SADD compilation time
$t_{\mathrm{SADD}}$. The number $N_{\mathrm{nr}}$ in parentheses counts
pulses that would not be possible if the original ion trajectories were
kept. Rows labeled \emph{mean/total} give arithmetic means of the ratios and
sums of the remaining quantities. The final row gives totals over all 20
circuits.

Although allowing trajectory changes gives the solver more options in each
local problem, Full SADD is not guaranteed to outperform pulse-only SADD on
every individual circuit. An accepted change also affects the opportunities
seen later, so the two chronological greedy passes can take different paths.
The circuit-resolved $\chi$ ratios in
\cref{tab:benchmark_resource_audit} show the resulting instance-to-instance
variation.

The total number of transport operations is not a direct measure of the
heating contribution used in our model. That contribution depends on how
much transport each ion undergoes before a two-qubit gate, as defined in
\cref{eq:heating_error}. The transport counts in
\cref{tab:benchmark_resource_audit} should therefore be read as a compiler
resource audit rather than as a substitute for the gate-level motional-error
exposure used in the noisy simulations.

\Cref{tab:benchmark_resource_audit} reports the phase-objective ratios and
infidelity ratios side by side for each circuit.
\Cref{fig:objective_fidelity_by_family} plots the two against each
other and resolves SADD's performance across circuit instances at the clearly
beneficial operating point $\delta\omega_0=10\,\mathrm{rad\,s^{-1}}$ and
$s_{\mathrm{ctrl}}=1$. Full SADD lowers both the phase objective and the
state infidelity on all twenty schedules, so the phase objective predicts the
direction of the effect without exception. The size of the fidelity gain is
only weakly predicted by the reduction in the objective. Over the corpus the
two improvements correlate at $+0.27$, with a bootstrap interval of
$[-0.16,+0.68]$ that includes zero. Separating the schedules by family shows
that the spread is not uniform across the corpus. The Ising schedules gain
more fidelity than their objective reduction alone suggests and the QPE
schedules gain less, so part of what looks like scatter in the pooled corpus
is a shift that follows circuit structure.

Part of that mismatch is control error, which $J_\Phi$ does not model at all.
Panel \textbf{(f)} shows this directly, with the correlation between the two
improvements rising as the control error is reduced and approaching $+0.64$
for error-free pulses. We attribute the remaining mismatch largely to how a
residual phase left at a critical point propagates through the circuit and
ultimately affects fidelity, which differs between circuits and is not
tracked by $J_\Phi$. This reflects a deliberate limitation of the objective,
since following the propagation exactly would require simulating the error
channel of the full circuit, the very computation the objective exists to avoid.

\section{Computational cost}
\label{appendix:complexity_parallelization}

The pass consists of two main tasks: processing the physical schedule and
solving a sequence of small local CP-SAT problems. For a schedule with $N$
ions and $T$ discrete time layers, replaying the ion trajectories,
toggling-frame signs, and phase proxy requires $O(NT)$ work. Finding the
control opportunities requires an additional scan of the processing-zone
occupation.

Consider now a control opportunity $u$ of duration $L_u$ that includes $N_u$
ions and $S_u$ represented sites. The variables describing the ion
trajectories, inserted pulses, pulse parity, and phase contributions give a
local model of size $O(N_uL_uS_u)$. On a general architecture, checking
possible motion between pairs of sites can require up to
$O(N_uL_uS_u^2)$ constraints. In the linear architecture studied here, each
site is connected only to its immediate neighbors. The number of allowed
moves therefore grows only linearly with $S_u$, and the corresponding local
model remains of size $O(N_uL_uS_u)$.

The implementation further restricts each opportunity to
$L_u\leq L_{\max}=16$ time layers and $N_u\leq N_{\max}=5$ ions. Thus,
within the architecture considered here, increasing the circuit size mainly
increases the number $M$ of local opportunities rather than the size of each
individual CP-SAT problem.

Furthermore, a time limit of $\tau=1\,\mathrm{s}$ is imposed for each local solve, such that the total
sequential solver time is bounded by $M\tau$. If the time limit is reached,
a candidate is kept only if it is feasible, passes the full-schedule replay,
and improves the phase objective. Otherwise, the original schedule is retained
for that opportunity. None of the Full-SADD opportunities in the benchmark
set reached the time limit.

The present implementation processes opportunities in chronological order,
because an accepted modification can change the schedule seen by later
local problems. This does not, however, require all opportunities to be
solved sequentially. Two opportunities are independent if they do not use
the same ions or control resources at overlapping times and do not affect
the same critical window. These dependencies can be represented by a
conflict graph, where each opportunity is a vertex, and two vertices are connected
when the corresponding opportunities cannot be solved at the same time.
Coloring this graph divides the set of opportunities into independent batches that
can be processed in parallel. The associated scaling can remain favorable as schedules grow, when a larger
schedule contains more local opportunities without making each
opportunity interact with an increasing fraction of the schedule. If the
bounds on the local ion number and time span remain fixed, and if critical
windows do not grow so long that they connect an increasing number of
opportunities, then each opportunity conflicts with only a bounded number
of others. Under this weak-scaling assumption, the conflict graph can then be colored using a bounded number of batches, independent of the total number of opportunities. The otherwise expensive volume of local problems in a larger schedule can then
be distributed across additional processors, leaving the CP-SAT part with
an approximately constant parallel depth. Such a batching scheme was not implemented or tested in the present work and presents an avenue for further development.

\end{document}